# Soliton oscillations in the Zakharov-type system at arbitrary nonlinearity-dispersion ratio

**L.G. Blyakhman[a], E.M. Gromov[a], B.A. Malomed[b,c], and V.V. Tyutin[1a]**

[a]*National Research University Higher School of Economics, Nizhny Novgorod 603155, Russia*
[b]*Department of Physical Electronics, School of Electrical Engineering, Faculty of Engineering, Tel Aviv University, and center for Light-Matter Interaction, Tel Aviv 69978, Israel*
[c]*ITMO University, St. Petersburg 197101, Russia*

**Abstract**

The dynamics of two-component solitons with a small spatial displacement of the high-frequency (HF) component relative to the low-frequency (LF) one is investigated in the framework of the Zakharov-type system. In this system, the evolution of the HF field is governed by a linear Schrödinger equation with the potential generated by the LF field, while the LF field is governed by a Korteweg-de Vries (KdV) equation with an arbitrary dispersion-nonlinearity ratio and a quadratic term accounting for the HF feedback on the LF field. The oscillation frequency of the soliton's HF component relative to the LF one is found analytically. It is shown that the solitons are stable against small perturbations. The analytical results are confirmed by numerical simulations.

**Highlights:**
>The dynamics of the two-component (HF-LF) soliton with initial displacement of the HF component is investigated.>The study is performed in the framework of the Zakharov-type system of coupled linear-Schrödinger and KdV equations, with an arbitrary ratio of nonlinearity and dispersion coefficients. >Analytical and numerical results are produced. > The oscillation frequency of the displaced HF component of the two-component soliton is found. > Stability of the perturbed two-component solitons is demonstrated>.

## 1. Introduction

Soliton solutions play a well-known role, predicting robust self-trapped eigenmodes in various models of wave propagation in nonlinear media with dispersion: surface waves on deep or shallow water, internal waves in a stratified fluid; electromagnetic, Langmuir and ion-acoustic waves in plasmas, optical pulses and beams in optical waveguides, matter waves in Bose-Einstein condensates, etc.[1-9].

The dynamics of intense high-frequency (HF) waves, such as surface waves on deep water and electromagnetic or Langmuir waves in plasmas, is self-consistently described in the second approximation of the theory of weakly nonlinear waves. The ubiquitous model appearing in this context is the nonlinear Schrödinger equation [10, 11], which produces envelope solitons as a result of the balance between the linear dispersion and effective cubic nonlinearity. A ubiquitous model equation for intense low-frequency (LF) waves, which may represent internal waves in a stratified fluid, surface waves on shallow water, and ion-acoustic waves in plasmas, amounts to the Korteweg - de Vries (KdV) equation, which gives rise to real solitons, supported by the balance between the quadratic nonlinearity and dispersion of LF waves.

---

[1]Corresponding author: Tel. +7(831)416-95-40.
*E-mail:*vtyutin@hse.ru (V.V. Tyutin).
Mailing address: RUSSIA, 603155 Nizhny Novgorod, 25/12 BolshayaPecherskaya St.

The co-propagation of HF and LF waves is governed by Zakharov-type systems of equations, in which the HF wave field is described by a linear Schrödinger equation with an effective potential generated by the LF wave field [12-14]. For electromagnetic and Langmuir waves in plasmas, this potential corresponds to perturbations in the plasma density caused by ion-acoustic waves. For surface waves, the effective potential corresponds to the surface flow produced by internal waves in the stratified fluid. If the HF and LF waves are traveling in the same direction, the intense LF wave field is described by the KdV equation [15-17], which includes the quadratic nonlinearity, dispersion, and the feedback of the HF field on the LF component in the form of a quadratic striction term.

Solitons in such a Zakharov system (ZS) were studied for the following cases: (i) for equal nonlinearity and dispersion coefficients of the LF wave ($\beta$ and $\gamma$, respectively, see Sec. 2 below) ($\gamma/\beta = 1$), a two-component Davydov-Scott (DS) soliton was found [18, 19]; oscillations of the HF component in which were investigated in [20]; (ii) for $\gamma/\beta \neq 1$, an asymptotic solution in the form of a multihump soliton, assuming a sufficiently small HF field amplitude in comparison to the LF one, was found in [14]; (iii) a localized steady state of the HF wave field was found, neglecting strictional action of the HF field on LF waves, given that there exists a LF solution in the form of a "depression" soliton [17]; (iv) new two-component solitons with two free parameters were found analytically in [21] in the asymptotic approximation for an arbitrary ratio of nonlinearity $\beta$ and dispersion $\gamma$; and, finally, a one-parameter family of exact two-component solitons under the condition of relatively weak dispersion, $\gamma < 3\beta$, was found in [21].

In the present work we aim to consider intra-soliton excitations, in the form of oscillations of the HF component about the LF component of the soliton, following work [21]. The study is performed in the framework of the ZS, consisting of a linear-Schrödinger and KdV equations with an arbitrary ratio of $\beta$ and $\gamma$ coefficients. The oscillation frequency of the HF component against the LF one is found in an integral form. The reported results demonstrate that the solitons found in [21] are stable. The analytical results are confirmed by systematic numerical simulations.

## 2. The coupled linear-Schrödinger – KdV system

We consider the co-propagation of the real LF field, $n(x,t)$, and a slowly varying envelope $U(x,t)$ of the complex HF field, $U(x,t)\exp(ik_0 x - i\omega_0 t)$, with the carrier wavenumber $k_0$ and frequency $\omega_0$. In geophysical hydrodynamics, the corresponding ZS system includes the linear Schrödinger equation for the surface waves and the KdV equation for internal waves, that are coupled by quadratic terms [22, 23]. The HF-LF interaction is resonantly enhanced in the case of the group synchronism, when the group velocity of the HF waves, $V_{\text{HF}} = (\partial \omega_{\text{HF}}/\partial k)_{k_0}$, coincides with the LF phase velocity $V_{\text{LF}} = (\partial \omega_{\text{LF}}/\partial k_{\text{LF}})$: $V_{\text{HF}} = V_{\text{LF}} \equiv V_{\text{SYN}}$. In this case, the system of the Schrödinger and KdV equations takes the following form, in the reference frame moving at group velocity $V_{\text{SYN}}$ (with $\xi = x - V_{\text{SYN}} t$) [17]:

$$2i\frac{\partial U}{\partial t} - \frac{\partial^2 U}{\partial \xi^2} + 2nU = 0, \qquad (1)$$

$$2V_{\text{SIN}}\frac{\partial n}{\partial t} - 6\beta \frac{\partial(n^2)}{\partial \xi} + \gamma \frac{\partial^3 n}{\partial \xi^3} = -\frac{\partial(|U|^2)}{\partial \xi}, \qquad (2)$$

where $\beta$ and $\gamma$ are the LF nonlinearity and dispersion coefficients. Then, under condition $\gamma/\beta = 1$, Eqs. (1)-(2) admit a two-component DS soliton solution [18, 19]:

$$U = A\operatorname{sech}\left(\frac{\xi - Vt}{\Delta}\right)\exp\left(-i\frac{t}{\Delta^2} - iV\xi\right), \quad n = -\frac{1}{\Delta^2}\operatorname{sech}^2\left(\frac{\xi - Vt}{\Delta}\right) < 0, \tag{3}$$

where $\Delta$ is an arbitrary width of the HF component of the soliton, and $V = (4\gamma/\Delta^2 - A^2\Delta^2)/(2V_{\text{SIN}})$ is its soliton's velocity. For a perturbed state of this soliton, oscillations of the HF component with respect to its LF counterpart, originally displayed in [20].

With an arbitrary value of the $\gamma/\beta$ ratio, the ZS based on Eqs. (1) and (2) admits an asymptotic solution [21],

$$U = A\operatorname{sech}^k\left(\frac{\xi - Vt}{\Delta}\right)\exp(-iV\xi - i\Omega t), \quad n = -\frac{\gamma}{\beta\Delta^2}\operatorname{sech}^2\left(\frac{\xi - Vt}{\Delta}\right),$$
$$k = (\sqrt{1 + 8\gamma/\beta} - 1)/2, \tag{4}$$

with the soliton's velocity and carrier frequency of the HF component given by

$$V = \frac{2\gamma}{\Delta^2}, \Omega = (2/\Delta^2)(1 - \gamma^2/\Delta^2). \tag{5}$$

For $\gamma \leq 3\beta$, the exact soliton solution of the system of equations (1)-(2) was also found in [21]:

$$U = \frac{\sqrt{18(3\beta - \gamma)}}{\Delta^2}\operatorname{sech}^2\left(\frac{\xi - Vt}{\Delta}\right)\exp(-iV\xi - i\Omega t), \quad n = -\frac{3}{\Delta^2}\operatorname{sech}^2\left(\frac{\xi - Vt}{\Delta}\right). \tag{6}$$

Below, the oscillations of the HF component of the soliton (4), with respect to the LF component, are studied analytically in the framework of the ZS based on Eqs. (1)-(2), for generic values of ratio $\gamma/\beta$. It is shown that the oscillation frequency grows as a function of $\gamma/\beta$.

## 3. Oscillations of a perturbed soliton

The system of equations (1)-(2) with zero boundary conditions at infinity, $(n, U)|_{\xi \to \pm\infty} \to 0$, gives rise to the following evolutions for integral moments, which may be used for an approximate analysis:

$$\frac{dN}{dt} \equiv \frac{d}{dt}\int_{-\infty}^{+\infty}|U|^2\,d\xi = 0, \tag{7}$$

$$\frac{d}{dt}\int_{-\infty}^{+\infty}K|U|^2\,d\xi = -\int_{-\infty}^{+\infty}n\frac{\partial(|U|^2)}{\partial\xi}\,d\xi, \tag{8}$$

$$\frac{d}{dt}\int_{-\infty}^{+\infty}\xi|U|^2\,d\xi = -\int_{-\infty}^{+\infty}K|U|^2\,d\xi, \tag{9}$$

where $U \equiv |U|\exp(i\varphi)$, $K \equiv \partial\varphi/\partial\xi$. The differentiation of (9) and substitution of Eq. (8) into the obtained expression yields an expression for the acceleration of the center of mass of the HF component:

$$\frac{d^2\bar{\xi}}{dt^2} = \frac{1}{N}\int_{-\infty}^{+\infty}n\frac{\partial(|U|^2)}{\partial\xi}\,d\xi, \tag{10}$$

where $\bar{\xi} = N^{-1}\int_{-\infty}^{+\infty}\xi|U|^2\,d\xi$ is the coordinate of the center of mass of the HF component. Assuming that the initially displayed HF component performed shuttle oscillations, keeping the soliton-like shape, i.e., $|U(\xi, \bar{\xi}, t)|^2 \equiv |U(\xi - Vt - \bar{\xi})|^2$, with a small displacement in comparison to the width of the HF component, $|\bar{\xi}| << \Delta$, we can use an asymptotic expansion of the density profile of the HF component, $|U(\xi - Vt - \bar{\xi})|^2$, in terms of $\bar{\xi}$:

$$|U(\xi - Vt - \bar{\xi})|^2 \approx |U(\xi - Vt)|^2 - \bar{\xi}\frac{\partial(|U(\xi - Vt)|^2)}{\partial\xi}. \tag{11}$$

The substitution of (11) into (10) yields an evolution equation for the coordinate of the center of mass of the HF component:

$$\frac{d^2\bar{\xi}}{dt^2} - \bar{\xi}\frac{1}{N}\int_{-\infty}^{+\infty}\frac{\partial n}{\partial\xi}\frac{\partial(|U(\bar{\xi}=0)|^2)}{\partial\xi}\,d\xi = 0, \tag{12}$$

where $U(\bar{\xi} = 0)$ is the unperturbed HF component of the soliton. In the case of a negative sign of the LF soliton component ($n < 0$), Eq. (12) takes form the harmonic-oscillator equation,

$$\frac{d^2\bar{\xi}}{dt^2} + \omega^2\bar{\xi} = 0, \tag{13}$$

where the squared oscillation frequency is

$$\omega^2 = -\frac{1}{N}\int_{-\infty}^{+\infty}\frac{\partial n}{\partial \xi}\frac{\partial\left(|U(\bar{\xi}=0)|^2\right)}{\partial \xi}d\xi > 0. \tag{14}$$

The dependence of $\omega\Delta^2$ on parameter of nonlinearity-to-dispersion ratio, $\gamma/\beta$, for the two-component soliton (4) is plotted in fig.1. For $\gamma/\beta = 1$, the oscillation frequency from (14) is $\omega = 2\sqrt{2/15}/\Delta^2$ and coincides with the frequency of oscillations of the HF component in the DS (3), which was obtained earlier in [20]. For $\gamma/\beta = 3$, the oscillation frequency from (14), $\omega = 4\sqrt{6/35}/\Delta^2$, corresponds to the HF oscillation frequency in the soliton (6). The persistence of such oscillations implies stability of the solitons (4) and (6).

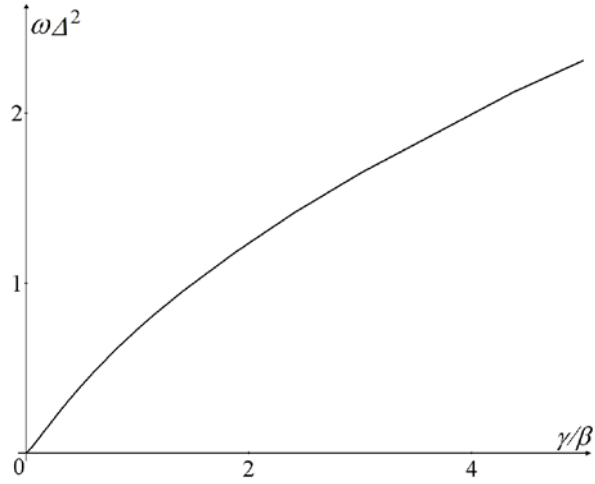

Fig.1. The curve for $\omega\Delta^2$ *versus* parameter $\gamma/\beta$, (see Eq. (4)), as obtained from Eq. (14).

### 4. Numerical results

The dynamics of the wave packets in the framework of the ZS based on Eqs. (1)-(2) for the coinciding group and phase velocities of the HF and LF waves, $V_{SYN} = 1$, was simulated with the input corresponding to a perturbed soliton (4), with the initially displaced HF component:

$$U(\xi, 0) = A\text{sech}^k\left(\frac{\xi-\xi_0}{\Delta}\right)\exp(-iV\xi), \quad n(\xi, 0) = -\frac{\gamma}{\beta\Delta^2}\text{sech}^2\left(\frac{\xi}{\Delta}\right), \tag{15}$$

where $V = 2\gamma/\Delta^2$, $k$ is the same as in Eq. (4), and $\xi_0$ is the initial displacement of the HF component relative to its LF counterpart. Figure 2 illustrates the resulting spatio-temporal dynamics of fields $|U(\xi, t)|$ (a) and $n(\xi, t)$ (b) for

$$\gamma = 2.5, \beta = 1.5 \ (\gamma/\beta = 5/3), \Delta = 1, \tag{16}$$

and $\xi_0 = 0.5$. The initial displacement of the HF component relative to the LF field leads to the shuttle oscillations of the former component about the latter one. In particular, at values of the parameters given by Eq. (16), the oscillation frequency, $\omega \approx 1.1$, is almost identical to that produced by Eqs. (14) and (4). The robust oscillatory motion of the displaced HF component implies dynamical stability of soliton (4) with respect to small perturbations. Similar numerical results, which clearly corroborate the stability of the solitons, were obtained for other values of the basic control parameter, $\gamma/\beta$.

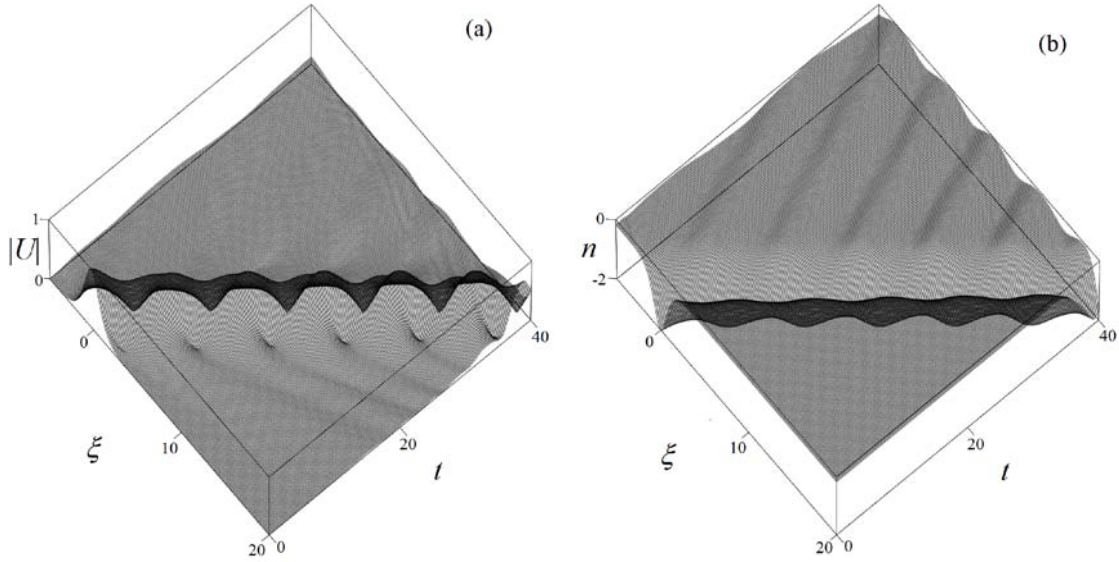

Fig. 2. A typical results produced by simulations of Eqs. (1)-(2) the spatio-temporal evolution of fields $|U(\xi,t)|$ (a) and $n(\xi,t)$ (b) for the input (15).

## 5. Conclusion

In the framework of the ZS (Zakharov's system), composed of coupled equations for interacting HF and LF (high- and low-frequency) fields, we have investigated internal dynamics of two-component solitons, initiated by a displacement of the HF component against the LF one. In the system, the HF field is governed by the linear Schrödinger equation with the effective spatio-temporal potential induced by the LF field. In turn, the LF field is described by the KdV equation with generic values of the ratio of the nonlinearity $\beta$ and dispersion $\gamma$ coefficients, while the feedback of the HF field is accounted for by a quadratic term added to the KdV equation. We have shown that a small initial displacement of the soliton's HF component gives rise to its persistent shuttle oscillations near the LF component with a constant frequency. This fact implies the dynamical stability of the two-component solitons in the ZS. The oscillation frequency has been found analytically in an integral form. The analytical results have been confirmed by numerical simulations of the solitons' dynamics.